\documentclass[aps,preprint,twocolumn,10pt]{revtex4}
 \pdfoutput=1
\usepackage{hyperref}
\usepackage{amsfonts}
\usepackage{amsmath}

\usepackage{amssymb}
\usepackage{graphicx}%
\usepackage{bigints}
\usepackage{graphicx}
\usepackage{subfigure}
\usepackage{epsf}
\usepackage{bm}
\usepackage{amsmath}
\usepackage{amsfonts}
\usepackage{amssymb}
\usepackage{graphicx}
\usepackage{tabularx}
\usepackage{color}%
\usepackage{multirow}
\providecommand{\U}[1]{\protect\rule{.1in}{.1in}}

\newcommand{\be}{\begin{equation}}
\newcommand{\ee}{\end{equation}}

\newcommand{\mincir}{\raise
-3.truept\hbox{\rlap{\hbox{$\sim$}}\raise4.truept\hbox{$<$}\ }}
\newcommand{\magcir}{\raise
-3.truept\hbox{\rlap{\hbox{$\sim$}}\raise4.truept\hbox{$>$}\ }}

\newcolumntype{Y}{>{\centering\arraybackslash}X}
\providecommand{\U}[1]{\protect\rule{.1in}{.1in}}

\usepackage{tikz,xcolor,hyperref}

\definecolor{lime}{HTML}{A6CE39}
\DeclareRobustCommand{\orcidicon}{%
	\begin{tikzpicture}
	\draw[lime, fill=lime] (0,0) 
	circle [radius=0.16] 
	node[white] {{\fontfamily{qag}\selectfont \tiny ID}};
	\draw[white, fill=white] (-0.0625,0.095) 
	circle [radius=0.007];
	\end{tikzpicture}
	\hspace{-2mm}
}

\foreach \x in {A, ..., Z}{%
	\expandafter\xdef\csname orcid\x\endcsname{\noexpand\href{https://orcid.org/\csname orcidauthor\x\endcsname}{\noexpand\orcidicon}}
}

\begin{document}

\title{Observational Constraints on Yukawa Cosmology and Connection with Black Hole Shadows}
\author{Esteban Gonzalez$^1$}
\email{esteban.gonzalez@ucn.cl}
\author{Kimet Jusufi$^2$}
\email{kimet.jusufi@unite.edu.mk}
\author{Genly Leon$^{3,4}$}
\email{genly.leon@ucn.cl (corresponding author)}
\author{Emmanuel N. Saridakis$^{3,5,6}$}
\email{msaridak@noa.gr}
\affiliation{$^1$Departamento de Física, Universidad Católica del Norte, Avenida Angamos 0610, Casilla 1280, Antofagasta, Chile}
\affiliation{$^2$Physics Department, State University of Tetovo, 
Ilinden Street nn, 1200, Tetovo, North Macedonia}
\affiliation{$^3$Departamento de Matem\'{a}ticas, Universidad Cat\'{o}lica del Norte, Avda.
Angamos 0610, Casilla 1280 Antofagasta, Chile}
\affiliation{$^4$Institute of Systems Science, Durban University of Technology, PO Box 1334,
Durban 4000, South Africa}

 \affiliation{$^5$National Observatory of Athens, Lofos Nymfon, 11852 Athens, 
Greece}

 \affiliation{$^6$CAS Key Laboratory for Researches in Galaxies and Cosmology, 
Department of Astronomy, University of Science and Technology of China, Hefei, 
Anhui 230026, P.R. China}

\begin{abstract}

We confront Yukawa modified cosmology, proposed in \href{https://arxiv.org/abs/2304.11492}{[Jusufi et al. arXiv:2304.11492]},  with data from  Supernovae Type Ia (SNe Ia)  and  Hubble parameter   (OHD) observations. Yukawa cosmology is obtained from  a Yukawa-like gravitational potential, with coupling parameter $\alpha$ and wavelength parameter $\lambda$, which gives rise to modified Friedmann equations. We show that the agreement with observations is very efficient, and within   $1\sigma$ confidence level  we find the best-fit parameters $\lambda=\left(2693_{-1262}^{+1191}\right)\, \rm Mpc$, $\alpha=0.416_{-0.326}^{+1.137}$, and a graviton mass of $m_{g}=\left(2.374_{-0.728}^{+2.095}\right)\times 10^{-42}\, \text{GeV}$. Additionally, we establish a connection between the effective dark matter and dark energy density parameters and the angular radius of the black hole shadow of the SgrA and M87 black holes in the low-redshift limit, consistent with the Event Horizon Telescope findings.  
\end{abstract}
\maketitle
\section{Introduction}

According to modern cosmology, the universe's large-scale structure is 
homogeneous and isotropic. Additionally, it is believed that cold dark 
matter, a type of matter that is not visible and only interacts through gravity, 
exists  \citep{Primack:1983xj, Peebles:1984zz, Bond:1984fp, Trimble:1987ee, 
Turner:1991id}. However, despite numerous efforts, there have not been any
direct detection of dark matter particles and their existence is only inferred 
from its gravitational effects on galaxies and larger structures. On the 
other hand, dark energy  is also introduced to 
explain the universe accelerated expansion \citep{Carroll:1991mt}, 
  supported by  numerous observations  
\citep{SupernovaCosmologyProject:1997zqe, SupernovaSearchTeam:1998fmf,  
SupernovaCosmologyProject:1998vns}.

The $\Lambda$CDM paradigm has proven 
to be the most successful model in modern cosmology. This scenario can describe 
cosmological observations with the least number of parameters 
\citep{Planck:2018vyg}. However, specific fundamental physics concepts remain to 
be fully understood, such as the microphysical nature of dark matter and dark 
energy. Since scalar fields play a significant role in the physical 
description of the universe in the inflationary scenario \citep{Guth:1980zm}, 
a quintessence scalar field is used in a 
generalization of the $\Lambda$CDM model \cite{Ratra:1987rm, Parsons:1995kt, 
Rubano:2001xi, Saridakis:2008fy,  Cai:2009zp,  WaliHossain:2014usl,
Barrow:2016qkh}.  Additionally, multi-scalar field models can describe various 
epochs of the cosmological 
history \citep{Elizalde:2004mq, Elizalde:2008yf, Skugoreva:2014ena, Saridakis:2016mjd, Paliathanasis:2018vru, Banerjee:2022ynv,
Santos:2023eqp}. Moreover, a unified description of the matter and dark energy 
epochs was presented for a class of scalar-torsion theories, providing a 
Hamiltonian description \citep{Leon:2022oyy}.
Nevertheless, there is direction in the literature 
that deviates from this line of thought and supports the idea that   
observations can be explained by altering Einstein's equations, leading to 
modified theories of gravity \citep{CANTATA:2021ktz,Leon:2009rc, DeFelice:2010aj, 
Nojiri:2010wj, Clifton:2011jh, Capozziello:2011et, DeFelice:2011bh, Xu:2012jf, 
Bamba:2012cp, Leon:2012mt, Kofinas:2014aka, Bahamonde:2015zma, Momeni:2015uwx, 
Cai:2015emx, Krssak:2018ywd,Dehghani:2023yph}. 

Concerning  dark matter, which is needed to explain the galaxy 
rotation curves \citep{Salucci:2018eie}, one of the first theories 
suggesting an explanation for the flatness of rotation curves was the Modified 
Newtonian dynamics (MOND)  proposed by Milgrom \citep{Milgrom:1983ca}, which
modifies Newton's law 
\citep{Ferreira:2009eg, Milgrom:2003, Tiret:2007kq, Kroupa:2010hf, Cardone:2010ru, 
Richtler:2011zk, Bekenstein:2004ne}. Other interesting proposals are the superfluid 
dark matter  \citep{Berezhiani:2015bqa},  the Bose-Einstein condensate 
\citep{Boehmer:2007um}, etc.

On the other hand, black holes are intriguing astronomical objects and  can 
potentially test the theories of gravity in strong gravitational fields.   One 
of the most fascinating aspects of black holes is their shadow image. The black 
hole's silhouette is a dark region that results from the immense gravitational 
pull of the black hole, which bends the path of light rays near it. 
Specifically, photons emitted from a bright source close to a black hole can 
either be drawn into the black hole or scattered away from it and into infinity. 
Additionally, critical geodesics separate the first two sets, known as unstable 
spherical orbits. By observing the critical photon geodesic trajectories in the 
sky, we can obtain the black hole shadow \citep{Takahashi:2004xh, Hioki:2009na, 
Brito:2015oca, Cunha:2015yba, Ohgami:2015nra, Moffat:2015kva, 
Abdujabbarov:2016hnw, Cunha:2018acu, Mizuno:2018lxz, Tsukamoto:2017fxq, 
Psaltis:2018xkc, Amir:2018pcu, Gralla:2019xty, Bambi:2019tjh, Cunha:2019ikd, 
Khodadi:2020jij, Perlick:2021aok, Vagnozzi:2022moj, Saurabh:2020zqg, 
Jusufi:2020cpn, Tsupko:2019pzg, Escamilla-Rivera:2022mkc,Jusufi:2021fek}. With 
the recent results, the black hole shadow for the supermassive black holes M87 
and Sgr A was confirmed by the Event Horizon Telescope (EHT) collaboration 
\citep{EventHorizonTelescope:2019dse, EventHorizonTelescope:2020qrl, 
EventHorizonTelescope:2021srq, EventHorizonTelescope:2022wkp, 
EventHorizonTelescope:2022apq, EventHorizonTelescope:2022wok}.

In the present paper,  we   follow an approach motivated by cosmology and 
quantum field theories; we aim to study the dark sector   by 
introducing the Yukawa potential \citep{Garny:2015sjg, Arvanitaki:2016xds, 
Desmond:2018euk, Desmond:2018sdy, Tsai:2021irw}. We adopt the viewpoint of  Verlinde and consider gravity as an entropic force caused by the changes in the 
system's information \citep{Verlinde:2010hp}. Verlinde further argued that dark matter is an 
apparent effect, i.e., a consequence of the baryonic matter 
\citep{Verlinde:2016toy}. Furthermore, the corresponding entropic force was   
used in deriving the corrected Friedmann equations due to the minimal length, 
as recently studied in \citep{Jusufi:2022mir, Millano:2023ahb, Jusufi:2023ayv, Jusufi:2023xoa}. 

In particular, as it was recently shown in \citep{Jusufi:2023xoa},  dark matter  
can be explained by the coupling between baryonic matter   through a 
long-range force  via the Yukawa gravitational potential. 
This coupling is characterized by the coupling parameter $\alpha$, the 
wavelength parameter $\lambda$, and the Planck length   $l_0$. The 
modified Friedmann equations are derived using Verlinde's entropic force 
interpretation of gravity based on the holographic scenario and the 
equipartition law of energy. An equation connects the dark matter density, dark 
energy density, and baryonic matter density. It is worth noting that dark matter 
is not associated with a particle but is an apparent effect. Dark energy is 
related to graviton mass and $\alpha$, indicating that the cosmological constant 
can be viewed as a self-interaction effect between gravitons. The model 
parameters were estimated as $\lambda \simeq 10^3$ [Mpc] and $\alpha \in 
(0.0385,0.0450)$.  

In this work, we are interested in performing detailed observational tests of the cosmological scenario based on Yukawa potential.
 In particular, we wish to constrain the parameters of 
the model using Supernovae Type Ia (SNe 
Ia)  and  Hubble parameter   (OHD) observations.
 Additionally, we are interested in 
investigating the connection to black hole physics.   In particular, at 
low redshifts, one can use the angular radius of the black hole shadow to 
constrain the Hubble constant independently.  The manuscript is organized as 
follows. In Sect. \ref{II}  we review the Yukawa cosmological scenario, while 
in Sect. \ref{Sec:Constraints} we extract   observational constraints. In 
Sect. \ref{IV}  we study the relation between the modified Yukawa cosmology and 
black hole shadows, and in Sect. \ref{conclusions}  we comment on our findings.

\section{Yukawa Modified Cosmology}
\label{II}

In this section, we shall review the model that was recently studied in  
\citep{Jusufi:2023xoa}. The gravitational potential considered is modified   via 
the non-singular Yukawa-type gravitational potential 
\begin{equation}
\label{eq11a}
\Phi(r)=-\frac{G M m}{\sqrt{r^2+l_0^2}}\left(1+\alpha\,e^{-\frac{r}{\lambda}}\right)|_{r=R},
\end{equation}
with $l_0$ being a small quantity of Planck length order, i.e. $l_0 \sim 
10^{-34}$ cm and $\alpha>0$. Note that the wavelength of
massive graviton we have $\lambda=\frac{\hbar}{m_g c}>10^{20} \rm m $,
that leads to $m_g<10^{-64}$ kg for the graviton mass \citep{Visser:1997hd}.   If 
we use the relation $F=-\nabla \Phi(r)|_{r=R}$, and by neglecting the term 
$\alpha l_0^2/R^2 \rightarrow 0$ we get the modified Newton's law of 
gravitation 
\begin{equation}\label{F5}
F=-\frac{G M m}{R^2} 
\left[1+\alpha\,\left(\frac{R+\lambda+\frac{l_0^2}{R}}{\lambda}\right)e^{-\frac{
R}{\lambda}}\right]\left[1+\frac{l_0^2}{R^2}\right]^{-3/2}.
\end{equation}
We can further elaborate that such a force can be obtained if we modify the expression for the entropy by means of Verlinde's entropic force interpretation. In this theory,  when a test particle or
excitation moves apart from the holographic screen, the magnitude
of the entropic force on this body has the form \citep{Verlinde:2010hp}
$F\triangle x=T \triangle S$. Specifically, one can get Newton's law of gravitation via the entropy-area relationship $S=A/4$, however in general, one can modify the total expression for the entropy as $S=A/4+\mathcal{S}(A)$ \citep{Jusufi:2023xoa}. One uses the holographic scenario and the equipartition law of energy to get the expression for force. The entropy of the surface changes by $d S=\left[1/4+\partial \mathcal{S}/{\partial A}\right]d A$; at the same time, we can relate the area of the surface to the number of bytes according to $ A=QN,$
where $Q$ is a fundamental
constant, and $N$ is the number of bytes. We can now use the equipartition law
of energy, we get the total energy on the surface via $E=Nk_B T/2$, and further taking $\triangle N=1$, and $\triangle A=Q$, we get the modified gravity force
\begin{equation}\label{F3}
F=-\frac{GMm}{R^2}\left(\frac{Q}{2\pi k_B
\eta}\right)\left[\frac{1}{4}+\frac{\partial \mathcal{S}}{\partial A}\right]_{A=4\pi R^2}.
\end{equation}
This means that, by changing the entropy, we end up with a modified law of gravity. Let us define $\eta=1/8\pi k_B$; we get $Q=1$,
\begin{equation}\label{F4}
F=-\frac{GMm}{R^2}\left[1+4\,\frac{\partial \mathcal{S}}{\partial A}\right]_{A=4\pi R^2}.
\end{equation}

Since for large-scale distances $l_0$ is unimportant, we can set it to zero, i.e. $l_0=0$. We now see that we can get the corrections
to the entropy if we compare 
\begin{equation}
1+\left(\frac{1}{2 \pi R}\right)\frac{dS}{dR}=\left[1+\alpha\,\left(\frac{R+\lambda}{\lambda}\right)e^{-\frac{R}{\lambda}}\right].
\end{equation}
Solving for entropy, we obtain \citep{Jusufi:2023xoa}
\begin{equation}\label{entropy1}
S=\pi R^2-2 \pi  \alpha\left(R^2+3\lambda R+3\lambda^2 \right) e^{-\frac{R}{\lambda}}.
\end{equation}
We have therefore shown that the Yukawa modified force follows from the modification of the entropy.  This method can be viewed as a generic result; if we modify the entropy, we modify Newton's law of gravity.  It is worth noting that the correction of the entropy can be viewed as a volume law entanglement to the entropy due to the gravitons. We see from the last equation that $\alpha$ appears only in the second term; hence we can say that $\alpha$ is a result of the entanglement due to the volume law entropy contribution. We proceed by studying the implications of the above-modified law of gravity in 
cosmology. First, we   assume the background spacetime to be spatially 
homogeneous and isotropic, described by the Friedmann-Robertson-Walker (FRW) 
metric
\begin{equation}
ds^2=-dt^2+a^2(t)\left[\frac{dr^2}{1-kr^2}+r^2(d\theta^2+\sin^2\theta
d\phi^2)\right],
\end{equation}
with  $R=a(t)r$,  $x^0=t, x^1=r$, and the two dimensional metric $ h_{\mu 
\nu}$, and where $k$  is the spatial curvature ($k = 0, 1, -1$
corresponding to flat, closed, and open universes, respectively). In addition, 
we have an apparent dynamic horizon, which the following relation can determine
$h^{\mu
\nu}(\partial_{\mu}R)\,(\partial_{\nu}R)=0$. 
It is easy to show that the apparent horizon radius for the FRW universe reads 
as
\begin{equation}
\label{radius}
 R=ar= {1}/{\sqrt{H^2+ {k}/{a^2}}}.
\end{equation}
On the other hand, we have a matter source which can be assumed to be a perfect
fluid described by the stress-energy tensor
\begin{equation}\label{T}
T_{\mu\nu}=(\rho+p)u_{\mu}u_{\nu}+pg_{\mu\nu},
\end{equation}
along with the continuity equation $\dot{\rho}+3H(\rho+p)=0$, with $H=\dot{a}/a$ 
being the Hubble parameter. 

Let us consider a compact spatial region $V$ with a compact boundary $\mathcal 
S$, corresponding to a sphere of radius $R= a(t)r$, where $r$ is a 
dimensionless quantity. Through Newton's law, we can write the gravitational force on a test particle $m$ near the surface 
\citep{Jusufi:2023xoa} as 
\begin{equation}\label{F6}
m\ddot{a}r=-\frac{GMm}{R^2}\left[1+\alpha\,\left(\frac{R+\lambda}{\lambda}\right)e^{-\frac{R}{\lambda}}\right] \left[1+\frac{l_0^2}{R^2}\right]^{-3/2}.
\end{equation}
In Newtonian cosmology, we can take $\rho=M/V$ inside the   volume 
$V=\frac{4}{3} \pi a^3 r^3$, hence we can  rewritten the above equation as 
\citep{Jusufi:2023xoa}
\begin{equation}\label{F7}
\frac{\ddot{a}}{a}=-\frac{4\pi G
}{3}\rho \left[1+\alpha\,\left(\frac{R+\lambda}{\lambda}\right)e^{-\frac{R}{\lambda}}\right] \left[1+\frac{l_0^2}{R^2}\right]^{-3/2},
\end{equation}
which is the dynamical equation for
Newtonian cosmology.  To obtain the Friedmann equations in general relativity, 
we must use the active gravitational mass $\mathcal M$ rather than the total 
mass $M$. By replacing $M$ with $\mathcal M$, we obtain
\begin{equation}\label{M1}
\mathcal M =-\ddot{a}
a^2r^3\left[1+\alpha\,\left(\frac{R+\lambda}{\lambda}\right)e^{-\frac{R}{\lambda}}\right] \left[1+\frac{l_0^2}{R^2}\right]^{-3/2},
\end{equation}
where the active gravitational mass can also be computed via
\begin{equation}\label{Int}
\mathcal M =2
\int_V{dV\left(T_{\mu\nu}-\frac{1}{2}Tg_{\mu\nu}\right)u^{\mu}u^{\nu}}.
\end{equation}
Using these equations, we obtain the modified acceleration equation for the dynamical
evolution of the  FRW universe \citep{Jusufi:2023xoa}
\begin{equation}\label{addot}
\frac{\ddot{a}}{a} =-\frac{4\pi G
}{3}(\rho+3p)\left[1+\alpha\,\left(\frac{R+\lambda} 
{\lambda}\right)e^{-\frac{R}{\lambda}}\right] 
\left[1-\frac{3\,l_0^2}{2\,R^2}\right].
\end{equation}
Furthermore, we can simplify the work since $l_0$ is a very small number; we can consider a series expansion around $x=1/\lambda$ via
\begin{eqnarray}
\left[1+\alpha\,\left(\frac{R+\lambda}{\lambda}\right)e^{-\frac{R}{\lambda}}
\right]=1+\alpha-\frac{1}{2}\frac{\alpha R^2}{\lambda^2}+ \cdots,
\end{eqnarray}
provided that $\alpha R^2/\lambda^2 \ll1$.  In general, we expect $\alpha <1$, 
and $\lambda $  to be some large number of magnitude comparable to the radius of 
the observable Universe $R \sim 10^{26}$ m. 

In summary, 
the corresponding 
Friedmann equation for $\alpha R^2/\lambda^2 \ll1$ becomes
\begin{equation}
\frac{\ddot{a}}{a}=- \frac{4 \pi G }{3}\sum_i 
\left(\rho_i+3p_i\right)\left[1+\alpha-\frac{1}{2}\frac{\alpha 
R^2}{\lambda^2}\right]\left[1-\frac{3 l_0^2}{2 R^2}\right], \label{Aaddot}
\end{equation}
where we have included several matter fluids with  a constant equation of state 
parameters $\omega_i$ along with the continuity equation $\dot{\rho}_i+3H(1+ 
\omega_i) \rho_i=0$, that yield an expression for densities $\rho_i=\rho_{i 0} 
a^{-3 (1+\omega_i)}$.  Inserting these into  \eqref{Aaddot}   and integrating 
we obtain \citep{Jusufi:2023xoa}
\begin{align}
d (\dot{a}^2+k)= & \frac{8\pi G}{3} \left[1+\alpha-\frac{1}{2}\frac{\alpha R^2}{\lambda^2}\right]\left[1-\frac{3\,l_0^2}{2\,R^2}\right] \nonumber \\
\times & d \left(\sum_i \rho_{i 0} a^{-1-3\omega_i)}\right).
\end{align} 
Using the fact that $R[a]= r a$, we 
further, get
\begin{align}\label{Fried1}
 \dot{a}^2+k = &   \frac{8\pi G}{3} \int \left[1+\alpha-\frac{1}{2}\frac{\alpha R[a]^2}{\lambda^2}\right]\left[1-\frac{3\,l_0^2}{2\,R[a]^2}\right] \nonumber \\
\times & \frac{d \left(\sum_i \rho_{i 0} a^{-1-3\omega_i)}\right)}{da} da,
\end{align}
with $r$ nearly a constant. Considering  the equations of state, $\omega_i 
\notin\{-1, 1/3\}$, we have 
 \begin{align}
  & \frac{\dot{a}^2}{a^2}+\frac{k}{a^2}=  \frac{8\pi  G }{3}  \left(\alpha  
\left(\frac{3 l_0^2}{4\lambda ^2}+1\right)+1\right) \sum_i \rho_{i0} a^{-3 
(1+\omega_{i})} \nonumber \\
  & -\frac{4 \pi  (\alpha +1) G l_0^2}{3 R^2} \sum_i\frac{ 3  \omega_{i}+1}{\omega_{i}+1}  \rho_{i0} 
a^{-3 (1+ \omega_{i})} \nonumber \\
   & +\frac{4 \pi  \alpha  G R^2}{3 \lambda ^2}   \sum_i \frac{ 1+ 3 
\omega_{i}}{1-3  \omega_{i}}  \rho_{i0}  a^{-3 (1+\omega _{i})},
 \end{align}
implying that at leading order terms ($l_0^2/\lambda ^2 \rightarrow 0$), 
\begin{align}
H^2+\frac{k}{a^2} = & \frac{8\pi G_{\rm eff}
}{3}\sum_i \rho_i  -\frac{1}{R^2}  \sum_i \Gamma_1(\omega_i)\rho_i  \nonumber 
\\& +\frac{4 \pi G_{\rm eff}}{3}R^2 \sum_{i}\Gamma_2(\omega_i)\rho_i, 
\label{Fried01}
\end{align}
where the Newton's constant is shifted as $G_{\rm eff}=G(1+\alpha)$,  along with 
the definitions \citep{Jusufi:2023xoa}
\begin{align}
\Gamma_1 (\omega_i ) & \equiv  \frac{4 \pi G_{\rm eff} l_0^2 }{ 3 }\left(\frac{1+3
\omega_i}{1+\omega_i}\right), \\
    \Gamma_2 (\omega_i )   & \equiv    \frac{\alpha\, (1+3\omega_i)}{  \lambda^2 
(1+\alpha) (1-3\omega_i)}.
\end{align}
 If, for example, we assume only a  matter source, at leading order terms  we 
can write
\begin{equation}
H^2+\frac{k}{a^2} =\frac{8\pi G_{\rm eff}
}{3}\rho-\frac{\Gamma_1}{R^2}\rho+\frac{4 \pi G_{\rm eff}}{3}\rho\,\Gamma_2 R^2.
\end{equation}
Focusing on  the flat case ($k=0$),  we have
$R^2=1/H^2$, yielding
\begin{equation}
H^2 \left(1+\Gamma_1 \rho\right)-\frac{4 \pi G_{\rm eff}}{3}\frac{\Gamma_2}{H^2}\rho=\frac{8\pi G_{\rm eff}
}{3}\rho.
\end{equation}
Finally, by expanding around $l_0$, making use of $ \left(1+\Gamma_1 
\rho\right)^{-1}\simeq \left(1-\Gamma_1 \rho\right)$, and neglecting the terms 
$\sim \mathcal{O}(l_0 \alpha^2/\lambda^2)$, we obtain 
 \begin{equation}\label{imeq}
H^2-\frac{4 \pi G_{\rm eff}}{3} \frac{\Gamma_2}{ H^2} \rho  =\frac{8\pi G_{\rm eff} }{3}\rho\left(1-\Gamma_1 \rho \right).
\end{equation}

\subsection{Late time universe}

Let us now study the modified Friedmann equation's phenomenological aspects. In particular, we are interested in studying the late 
universe, which implies we can neglect the quantum effects by setting $l_0 
\rightarrow 0$ [$\Gamma_1=0$]. This gives
\begin{equation}
H^2-\frac{4 \pi G_{\rm eff}}{3}\frac{\sum_i \Gamma_2(\omega_i)\,\rho_i}{H^{2}}=\frac{8\pi G_{\rm eff} }{3}\,\sum_i\rho_i,
\end{equation}
and using   $\rho_{\rm crit}=\frac{3}{8 \pi G}H_0^2$ 
we acquire two solutions:
\begin{align}\label{eq43}
  E^2&=\frac{(1+\alpha)}{2}\,\sum_i\Omega_i \notag \\
   &\pm  \frac{\sqrt{(\sum_i\Omega_i)^2 (1+\alpha)^2+2 \Gamma_2(\omega_i) \Omega_i (1+\alpha)/H_0^2}}{2},
\end{align}
where  $\Omega_i=\Omega_{i0}(1+z)^{3(1+\omega_i)}, \, \Omega_{i0}=  8 \pi G \rho_{i0}/(3H_0^2)$, with $E=H/H_{0}$.  In addition, we point out that the total quantity $\Omega^2$ in the square root should be viewed as the root-mean-square density energy, i.e $\Omega\equiv \sqrt{\left\langle \Omega^2 \right\rangle}$ along with $\left\langle \Omega^2 \right\rangle=\left\langle \Omega_B^2 \right\rangle+\left\langle \Omega^2_{\Lambda} \right\rangle$.
As  explained in \citep{Jusufi:2023xoa}, the most interesting implication of the 
last equation relies on the physical interpretation of the term $
     2  \Gamma_2(\omega_i) \Omega_i (1+\alpha)/H_0^2$. In particular,  it was 
shown that this term precisely mimics the effect of cold dark matter of the 
$\Lambda$CDM model. Taking the term $\Gamma_2 \Omega_i$ and set $\omega_i=0$ 
  we define the   quantity (here we shall add the constant term $c$ 
to make the equation consistent) \citep{Jusufi:2023xoa}
\begin{equation}
   \frac{\Omega^2_{D}(1+\alpha)^2}{{(1+z)^3}}\equiv \frac{2 \Gamma_2 
\Omega_i(1+\alpha)}{H_0^2}.  \label{eq30}
\end{equation}
Thus,  we can obtain an equation for dark matter as  
\begin{equation}\label{eqDM}
    \Omega_{D}= \frac{c}{\lambda H_0\,(1+\alpha)}\sqrt{2 \alpha \Omega_B} 
\,{(1+z)^{3} }.
\end{equation}
From this equation, we can deduce that dark matter 
can be viewed as an effective sector,   a manifestation of modified
Newton's law, quantified by   $\sim \alpha\, \Omega_B$. 
Additionally, we  define the   quantity  
\begin{equation}
    \Omega_{\Lambda}= \frac{c^2}{\lambda^2 H^2_0}\frac{\alpha}{(1+\alpha)^2}. \label{Eq.51}
\end{equation}
 Finally, comparing the last expression with  $\rho_{\Lambda} = \frac{\Lambda 
c^2}{8 \pi\, G}$,
we can estimate the effective cosmological constant to be  
$
\Lambda = \frac{3\,m_g^2 c^2\,\alpha}{\hbar^2\, (1+\alpha)^2 }$.
Note that one can   combine the above expressions and 
relate baryonic matter with the effective   dark matter  and  dark energy, 
acquiring   
\begin{equation}\label{DMCC}
   \Omega_D= \sqrt{2\, \Omega_B  \Omega_{\Lambda}}{(1+z)^3}.
\end{equation}

In summary,  Eq. \eqref{eq43} can be re-written as
\citep{Jusufi:2023xoa}
\begin{eqnarray}
   {E^2(z)}&=&(1+\alpha) \left[\Omega_B 
(1+z)^{3}+\Omega_{\Lambda}\right].
\end{eqnarray}

We can now introduce the split 
 $\Omega_B (1+z)^{3} \to \Omega^{\Lambda CDM}_B(1+z)^3+\Omega_D^{\Lambda CDM}$, 
hence, to get the $\Lambda$CDM-like model, we can write 
\begin{equation}\label{Eq.49}
{E^2(z)}= (1+\alpha) \left[\Omega^{\Lambda CDM}_B (1+z)^{3} 
+\Omega_D^{\Lambda CDM}+\Omega_{\Lambda}\right],
\end{equation}
where 
\begin{eqnarray}
    \Omega_D^{\Lambda CDM}= \frac{c}{\lambda H_0\,(1+\alpha)}\sqrt{2 \alpha 
\Omega^{\Lambda CDM}_B} \,{(1+z)^{3} }.
\end{eqnarray}

\section{Observational constraints}
\label{Sec:Constraints}

In the previous section, we presented Yukawa-modified cosmology
Hence, in this section, we can proceed to observational confrontation with
Hubble parameter data (OHD) and   Type Ia supernovae (SN Ia) data, in order 
to extract  constraints of the free parameters.  For this purpose, 
we compute the best-fit of the free parameters and their corresponding 
confidence regions at $1\sigma \,(68.3\%)$ of confidence level (CL) with the 
affine-invariant Markov chain Monte Carlo (MCMC) 
method~\citep{Goodman_Ensemble_2010}, implemented in the pure-Python code emcee 
\citep{Foreman-Mackey:2012any}. In particular, we have considered $100$ chains or 
``walkers'', using the autocorrelation time provided by the emcee module as a 
convergence test. In this sense, we computed at every $50$ step the 
autocorrelation time, $\tau_{\text{corr}}$, of the chains. If the current step 
is larger than $50\tau_{\text{corr}}$ and the values of $\tau_{\text{corr}}$ 
changed by less than $1\%$, then we will consider that the chains are converged, 
and the constraint is stopped. We discard the first $5\tau_{\text{corr}}$ steps 
as ``burn-in'' steps. Finally, we compute the mean acceptance fraction of the 
chains, which must have a value between $0.2$ and $0.5$ 
\citep{Foreman-Mackey:2012any} and can be modified by the stretch move provided by 
the emcee module.

For this Bayesian statistical analysis, we need to construct the following Gaussian likelihoods:
\begin{equation}\label{likelihoods}
\mathcal{L}_{\text{OHD}}\propto\exp{\left(-\frac{\chi_{\text{OHD}}^{2}}{2}\right)},\;
    \mathcal{L}_{\text{SNe}}\propto\exp{\left(-\frac{\chi_{\text{SNe}}^{2}}{2}\right)},
\end{equation}
where $\chi_{\text{OHD}}^{2}$ and $\chi_{\text{SNe}}^{2}$ are the merit function of the  OHD and SNe Ia data, respectively. The Gaussian likelihood for the joint analysis SNe Ia+OHD is constructed as $\mathcal{L}_{\text{joint}}=\mathcal{L}_{\text{SNe}}+\mathcal{L}_{\text{OHD}}$.

In the following subsections, we will briefly describe the construction of the merit function of each data set.

\subsection{\label{sec:OHD} Observational Hubble parameter data}

For the OHD, we consider the sample compiled by Magaña et al. 
\citep{Magana:2017nfs},  which consists of $51$ data points in the 
redshift range $0.07\leq z\leq 2.36$, and for which we construct their merit 
functions as
\begin{equation}\label{meritOHD}
    \chi_{\text{OHD}}^{2}=\sum_{i=1}^{51} 
{\left[\frac{H_{i}-H_{th}(z_{i},\theta)}{\sigma_{H,i}}\right]^{2}},
\end{equation}
where $H_{i}$ is the observational Hubble parameter at redshift $z_{i}$ with an associated error $\sigma_{H, i}$, all of them provided by the OHD sample, $H_{th}$ is the theoretical Hubble parameter at the same redshift, and $\theta$ encompasses the free parameters of the model under study. 

The theoretical Hubble parameter is obtained from Eq. \eqref{Eq.49},  which we 
conveniently rewrite as
\begin{small}
\begin{equation}\label{YukawaLCDMLike}
{E^{2}(z)}=\left(1+\alpha\right)\left[\left(\Omega_{B,0}+ 
\sqrt{2\Omega_{B,0}\Omega_{\Lambda,0}}\right)\left(1+z\right)^{3} +\Omega_{\Lambda,0}\right],
\end{equation}
\end{small}
with $\Omega_{\Lambda,0}$ given by Eq. \eqref{Eq.51}, while    the 
condition $H(z=0)=H_{0}$, leads to
\begin{equation}\label{alphaconstraint}
    1+\alpha=\left[{\Omega_{B,0}+\sqrt{2\Omega_{B,0}\Omega_{\Lambda,0}}+\Omega_{\Lambda,0}}\right]^{-1},
\end{equation}
and the free parameters of the Yukawa-modified cosmology are 
$\theta=\left\{H_{0};\Omega_{B,0};\Omega_{\Lambda,0}\right\}$.  Note that  one 
has the relations
\begin{eqnarray}
    \left(1+\alpha\right)\Omega_{B,0} &  \equiv\Omega_{B,0}^{\Lambda\text{CDM}}, 
\label{OB0def}\\
    \left(1+\alpha\right)\sqrt{2\Omega_{B,0} \Omega_{\Lambda,0}} & 
\equiv\Omega_{DM,0}^{\Lambda\text{CDM}}, \label{ODMdef}\\
    \left(1+\alpha\right)\Omega_{\Lambda,0}  & 
\equiv\Omega_{\Lambda,0}^{\Lambda\text{CDM}}, \label{OLdef}
\end{eqnarray}
and Eq. \eqref{YukawaLCDMLike} becomes
\begin{equation}\label{HLCDM}
   {E^{2}(z)}=\left(\Omega_{B,0}^{\Lambda\text{CDM}}+\Omega_{DM,0}^{\Lambda\text{CDM}}\right)\left(1+z\right)^{3}+\Omega_{\Lambda,0}^{\Lambda\text{CDM}},
\end{equation}
which is related to the Hubble parameter for the standard $\Lambda$CDM model through $H(z)=H_0 E(z)$, with  $H_
{0}=H_{0}^{\Lambda\text{CDM}}$.

It is important to mention that  the OHD, as well as the SNe Ia data, are 
not able to independently constraint $\Omega_{B,0}^{\Lambda\text{CDM}}$ and 
$\Omega_{DM,0}^{\Lambda\text{CDM}}$. Thus, for the $\Lambda$CDM scenario, we 
define 
$\Omega_{m,0}\equiv\Omega_{B,0}^{\Lambda\text{CDM}}+\Omega_{DM,0}^{\Lambda\text{ 
CDM}}$, where the condition $H(z=0)=H_{0}$ leads to 
$\Omega_{\Lambda,0}^{\Lambda\text{CDM}}=1-\Omega_{m,0}$, and the free parameters 
of the $\Lambda$CDM scenario are $\theta=\left\{H_{0};\Omega_{m,0}\right\}$. 
Therefore, we consider the following priors for our MCMC analysis:
$H_{0}=100\frac{km/s}{Mpc}\,h$, with $0.55<h<0.85$, $0<\Omega_{m,0}<1$, 
$0<\Omega_{B,0}<0.2$,   $0<\Omega_{\Lambda,0}<1$,  and the condition $\alpha>0$ 
implies 
$0<\Omega_{B,0}+\sqrt{2\Omega_{B,0}\Omega_{\Lambda,0}}+\Omega_{\Lambda,0}<1$.

\subsection{\label{sec:SNe} Type Ia supernovae data}

For the SNe Ia data, we consider the Pantheon+ sample \citep{Brout:2022vxf},  
which is the successor of the original Pantheon sample 
\citep{Pan-STARRS1:2017jku} and consist of 1701 data points in the redshift range 
$0.001\leq z\leq 2.26$. In this case, the merit function can be conveniently 
constructed in matrix notation (denoted by bold symbols) as
\begin{equation}\label{meritSNe}
    \chi_{\text{SNe}}^{2}=\mathbf{\Delta D}(z,\theta,M)^{\dagger}\mathbf{C}^{-1}\mathbf{\Delta D}(z,\theta,M),
\end{equation}
where $[\mathbf{\Delta D}(z,\theta,M)]_{i}= m_{B,i}-M-\mu_{th}(z_{i},\theta)$ 
and $\mathbf{C}=\mathbf{C}_{\text{stat}}+\mathbf{C}_{\text{sys}}$ is the total 
uncertainty covariance matrix, where the matrices $\textbf{C}_{\text{stat}}$ and 
$\textbf{C}_{\text{sys}}$ accounts for the statistical and systematic 
uncertainties, respectively. In this expression, $\mu_{i}=m_{B, i}-M$ is the 
observational distance modulus of Pantheon+ sample, obtained by a modified 
version of the Trip's formula \citep{Tripp:1997wt}, with three nuisance 
parameters calibrated to zero with the BBC (BEAMS with Bias Corrections) 
approach \citep{Kessler:2016uwi}. Therefore, the Pantheon+ sample provides 
directly the corrected apparent B-band magnitude $m_{B, i}$ of a fiducial SNe Ia 
at redshift $z_{i}$, with $M$ the fiducial magnitude of an SNe Ia, which must be 
jointly estimated with the free parameters of the model under study.

The theoretical  distance modulus for a spatially flat FLRW spacetime is given 
by
\begin{equation}\label{theoreticaldistance}
    \mu_{th}(z_{i},\theta)=5\log_{10}{\left[\frac{d_{L}(z_{i},\theta)}{\text{Mpc}}\right]}+25,
\end{equation}
with $d_{L}(z_{i},\theta)$ the  luminosity distance given by
\begin{equation}\label{luminosity}
    d_{L}(z_{i},\theta)=c(1+z_{i})\int_{0}^{z_{i}}{\frac{dz'}{H_{th}(z',\theta)}},
\end{equation}
where~$c$ is the speed  of light given in units of $\text{km/s}$. Note that the 
luminosity distance depends on the theoretical Hubble parameter, which is given 
by Eq. \eqref{YukawaLCDMLike} for the Yukawa cosmology and Eq. \eqref{HLCDM} for 
the $\Lambda$CDM model. Therefore, we only add to the free parameters the 
nuisance parameter $M$, for which we consider the following prior to our MCMC 
analysis: $-20<M<-18$.

Similarly to the Pantheon sample, there is a degeneration between the nuisance 
parameter $M$ and $H_{0}$. Hence, to constraint the free parameter $H_{0}$ using 
SNe Ia data alone, it is necessary to include the SH0ES (Supernovae and $H_{0}$ 
for the Equation of State of the dark energy program) Cepheid host distance anchors 
of the form
\begin{equation}\label{Cepheidmerit}
    \chi^{2}_{\text{Cepheid}}=\Delta\textbf{D}_{ 
\text{Cepheid}}\left(M\right)^{\dagger}\textbf{C}^{-1}\Delta\textbf{D}_{\text{
Cepheid}}\left(M\right),
\end{equation}
where 
$\left[\Delta\textbf{D}_{\text{Cepheid}}\left(M\right)\right]_{i}=\mu_{i}
\left(M\right)-\mu_{i}^{\text{Cepheid}}$,  with $\mu_{i}^{\text{Cepheid}}$ the 
Cepheid calibrated host-galaxy distance obtained by SH0ES \citep{Riess:2021jrx}. 
Hence, for the correspondence,  we use the Cepheid distances as the 
``theory'' instead of using the model under study to calibrate $M$, considering 
that the difference $\mu_{i}\left(M\right)-\mu_{i}^{\text{Cepheid}}$ is 
sensitive to $M$ and $H_{0}$ and also is largely insensitive to other parameters 
like $\Omega_{m,0}$. In this sense, the Pantheon+ sample provides 
$\mu_{i}^{\text{Cepheid}}$, and the total uncertainty covariance matrix for 
Cepheid is contained in the total uncertainty covariance matrix $\mathbf{C}$. 
Therefore, we can define the merit function for the SNe Ia data as
\begin{equation}\label{SNemeritfull}
    \chi_{\text{SNe}}^{2}=\mathbf{\Delta D'}(z,\theta,M)^{\dagger}\mathbf{C}^{-1}\mathbf{\Delta D'}(z,\theta,M),
\end{equation}
where
\begin{equation}\label{SNeresidual}
    \Delta\mathbf{D'}_{i}=\left\{\begin{array}{ll}
             m_{B,i}-M-\mu_{i}^{\text{Cepheid}} & i\in\text{Cepheid host} \\
             \\ m_{B,i}-M-\mu_{th}(z_{i},\theta) & \text{otherwise}
             \end{array}
   \right..
\end{equation}

It is essential to  mention that  from now on, we will omit the free parameter 
$M$ and we will focus our analysis only on the free parameters of each model. 
Besides, considering that the best-fit parameters minimize the merit function, 
we can use the evaluation of the best-fit parameters in the merit function, 
$\chi_{\text{min}}^{2}$, as an indicator of the goodness of the fit: the smaller 
the value of $\chi_{\text{min}}^{2}$ is, the better is the fit.
 

\subsection{\label{sec:Results} Results and discussions}

\begin{figure}
    \centering
    \includegraphics[scale = 0.35]{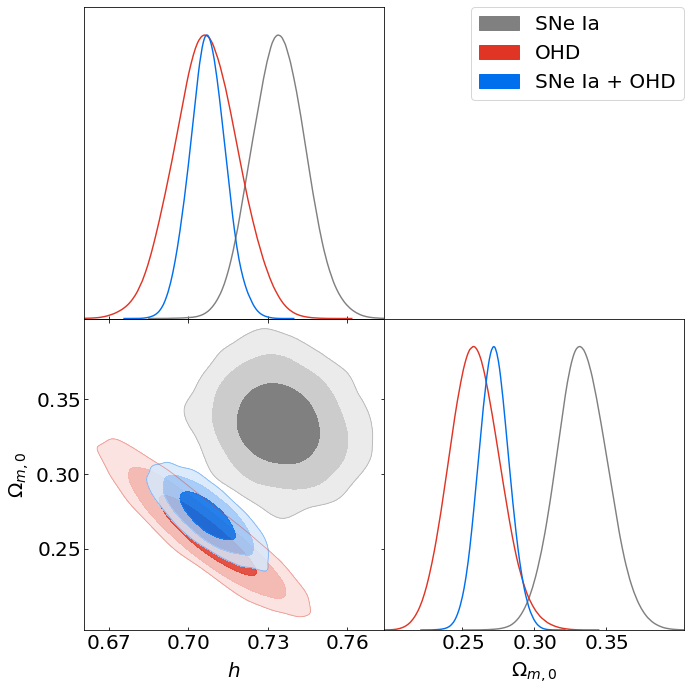}
    \caption{{\it{\label{fig:TriangleLCDM} Posterior 1D distribution and joint 
marginalized regions of the free parameters space of the $\Lambda$CDM model, 
obtained by the MCMC analysis described in Section \ref{Sec:Constraints}. The 
admissible joint regions correspond to $1\sigma$, $2\sigma$, and $3\sigma$ CL, 
respectively. The best-fit values for each model free parameter are shown in 
Table \ref{tab:best-fits}.}}}
\end{figure}

\begin{figure}
    \centering
    \includegraphics[scale = 0.35]{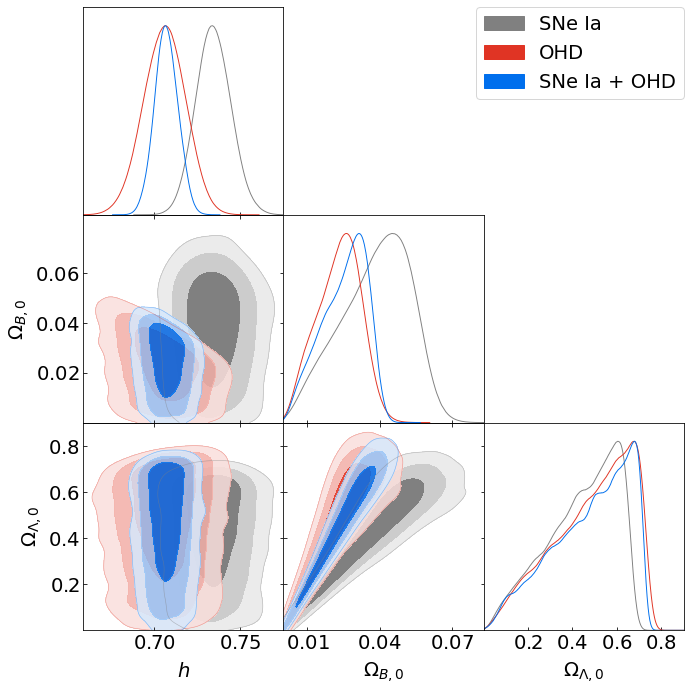}
    \caption{\label{fig:TriangleYukawa} {\it{Posterior 1D distribution and joint 
marginalized regions of the free parameters space of the Yukawa modified 
cosmology, obtained by the MCMC analysis described in Section 
\ref{Sec:Constraints}. The admissible joint regions correspond to $1\sigma$, 
$2\sigma$, and $3\sigma$ CL, respectively. The best-fit values for each model 
free parameter are shown in Table \ref{tab:best-fits}.}}}
\end{figure}

In Table \ref{tab:MCMC},  we present the total number of steps, the values used 
for the stretch move, the mean acceptance fraction, and the autocorrelation time 
for the free parameters $h$ and $\Omega_{m,0}$ of the $\Lambda$CDM model, and 
$h$, $\Omega_{B,0}$, and $\Omega_{\Lambda,0}$ of the Yukawa modified cosmology. 
Additionally, in Table \ref{tab:best-fits}, we present their respective 
best-fit values at $1\sigma$ CL with the corresponding $\chi_{\text{min}}^{2}$ 
criteria. In Figs. \ref{fig:TriangleLCDM} and \ref{fig:TriangleYukawa}, we 
depict the posterior 1D distribution and the joint marginalized regions of the 
free parameters space of the $\Lambda$CDM model and the Yukawa-modified 
cosmology. The admissible joint regions presented correspond to $1\sigma$, 
$2\sigma\,(95.5\%)$, and $3\sigma\,(99.7\%)$ CL, respectively. These results 
were obtained by the MCMC analysis described in Section \ref{Sec:Constraints} 
for the SNe Ia data, OHD, and their joint analysis.

\begin{table*}
    \centering
        \begin{tabularx}{\textwidth}{YYYYYYYY}
    \hline\hline
        \multirow{2}{*}{Data} & \multirow{2}{*}{Total Steps} & \multirow{2}{*}{$a$} & \multirow{2}{*}{MAF} & \multicolumn{4}{c}{$\tau_{\text{corr}}$} \\
        \cline{5-8}
         & & & & $h$ & $\Omega_{m,0}$ & $\Omega_{B,0}$ & $\Omega_{\Lambda,0}$ \\
         \hline
         \multicolumn{8}{c}{$\Lambda$CDM model} \\
         SNe Ia & $1250$ & $5.0$ & $0.35$ & $24.9$ & $23.6$ & $\cdots$ & $\cdots$ \\
         OHD & $2100$ & $5.0$ & $0.31$ & $31.9$ & $32.2$ & $\cdots$ & $\cdots$ \\
         SNe Ia+OHD & $1250$ & $5.0$ & $0.35$ & $24.0$ & $22.8$ & $\cdots$ & $\cdots$ \\
         \hline
         \multicolumn{8}{c}{Yukawa cosmology} \\
         SNe Ia & $2500$ & $3.0$ & $0.38$ & $38.1$ & $\cdots$ & $48.1$ & $49.8$ \\
         OHD & $4150$ & $2.5$ & $0.37$ & $64.1$ & $\cdots$ & $82.2$ & $80.7$ \\
         SNe Ia+OHD & $2300$ & $3.0$ & $0.38$ & $38.5$ & $\cdots$ & $45.8$ & $45.2$ \\
         \hline\hline
    \end{tabularx}
    \caption{ \label{tab:MCMC} The total number of steps, stretch move ($a$), 
mean acceptance fraction (MAF), and autocorrelation time ($\tau_{\text{corr}}$) 
for the free parameters space of the $\Lambda$CDM model and the Yukawa modified 
cosmology. The values were obtained when the convergence test described in 
Section \ref{Sec:Constraints} is fulfilled for an MCMC analysis with 100 chains 
and the flat priors $h\in F(0.55,0.85)$, $\Omega_{m,0}\in F(0,1)$, 
$\Omega_{B,0}\in F(0,0.2)$, and $\Omega_{\Lambda,0}\in F(0,1)$. Additionally, we
consider 
the constraint 
$0<\Omega_{B,0}+\sqrt{2\Omega_{B,0}\Omega_{\Lambda,0}}+\Omega_{\Lambda,0}<1$ 
for 
the Yukawa cosmology.}
\end{table*}

\begin{table*}
    \centering
       \begin{tabularx}{\textwidth}{YYYYYY}
        \hline\hline
        \multirow{2}{*}{Data} & \multicolumn{4}{c}{Best-Fit values} & \multirow{2}{*}{$\chi_{\text{min}}^{2}$} \\
         \cline{2-5}
         & $h$ & $\Omega_{m,0}$ & $\Omega_{B,0}$ & $\Omega_{\Lambda,0}$ & \\
        \hline
        \multicolumn{6}{c}{$\Lambda$CDM model} \\
        SNe Ia & $0.734\pm 0.010$ & $0.333\pm 0.018$ & $\cdots$ & $\cdots$ & $1523.0$ \\
        OHD & $0.707\pm 0.012$ & $0.259\pm 0.018$ & $\cdots$ & $\cdots$ & $27.5$ \\
        SNe Ia+OHD & $0.707\pm 0.007$ & $0.272\pm 0.011$ & $\cdots$ & $\cdots$ & $1576.7$ \\
        \hline
        \multicolumn{6}{c}{Yukawa cosmology} \\
        SNe Ia & $0.734\pm 0.010$ & $\cdots$ & $0.040_{-0.017}^{+0.013}$ & $0.468_{-0.206}^{+0.141}$ & $1523.0$ \\
        OHD & $0.706\pm 0.012$ & $\cdots$ & $0.024_{-0.011}^{+0.008}$ & $0.522_{-0.233}^{+0.153}$ & $27.5$ \\
        SNe Ia+OHD & $0.707\pm 0.007$ & $\cdots$ & $0.027_{-0.012}^{+0.008}$ & $0.513_{-0.229}^{+0.153}$ & $1576.7$ \\
        \hline\hline
    \end{tabularx}
     \caption{\label{tab:best-fits} Best-fit values and $\chi_{\text{min}}^{2}$ 
criteria for the $\Lambda$CDM model and the Yukawa cosmology. The 
values were obtained by the MCMC analysis  described in Section 
\ref{Sec:Constraints}, and the uncertainties presented correspond to 
$1\sigma$ CL.}
\end{table*}

As we can see from  Table \ref{tab:best-fits}, the values obtained for the 
$\chi^{2}_{\text{min}}$ criteria show that the Yukawa modified cosmology can  fit the observational data of SNe Ia, OHD and SNe Ia+OHD as accurately as the 
$\Lambda$CDM model. Even more, the value of the Hubble constant is the same in 
both models, which agrees with our previous identification in Eqs. 
\eqref{YukawaLCDMLike} and \eqref{HLCDM}, where 
$H_{0}=H_{0}^{\Lambda\textbf{CDM}}$. The only difference between these models 
relies on the rescaling of energy densities due to the contribution of the 
$\alpha$ parameter. On physical grounds, the main difference between these 
models are that in Yukawa cosmology, dark matter is effective and 
precisely mimics the cold dark matter of the $\Lambda$CDM scenario.

\begin{figure}
    \centering
    \includegraphics[scale = 0.56]{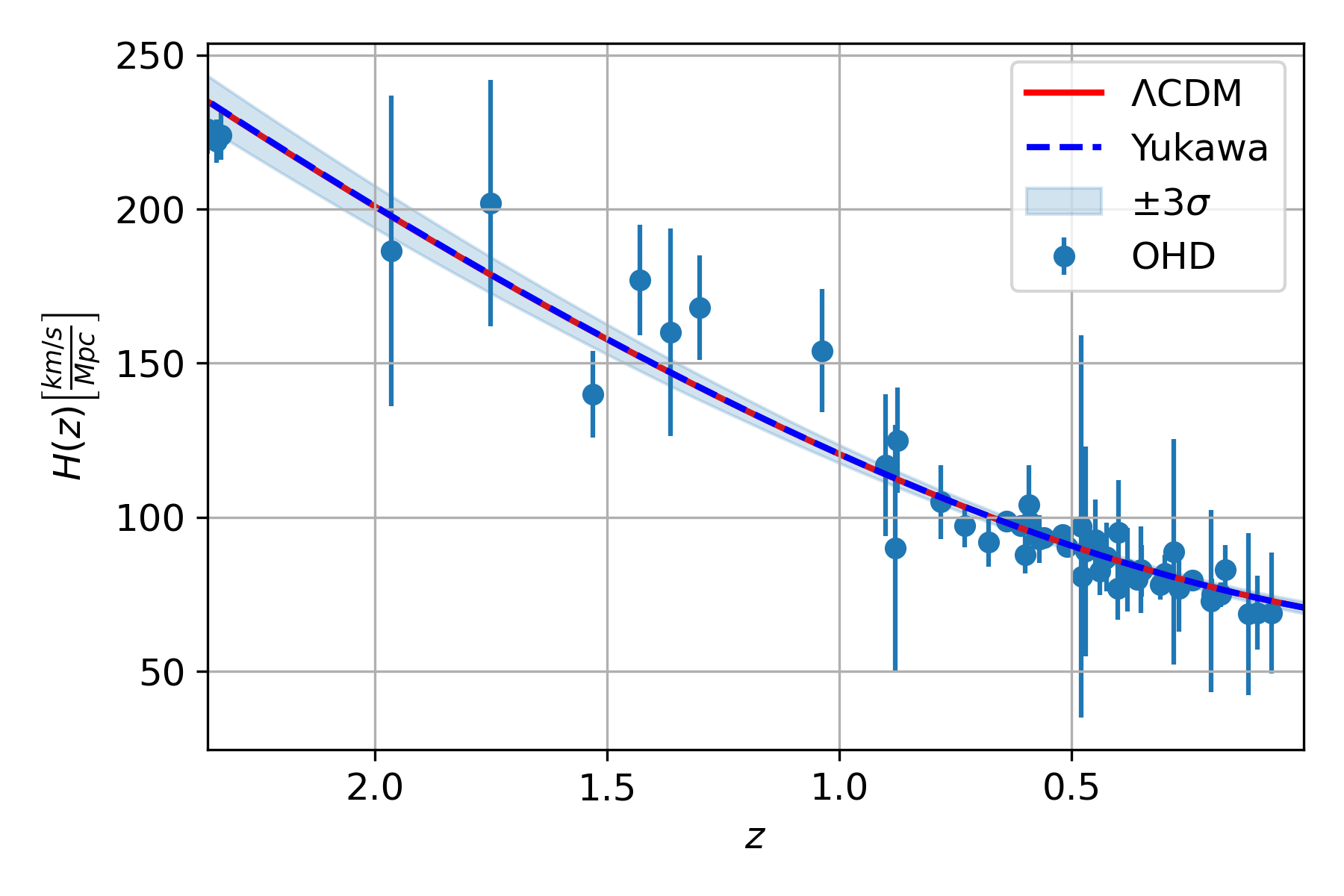}
    \caption{{\it{\label{fig:HYukawaplot} Hubble parameter as a function of 
redshift for the $\Lambda$CDM and Yukawa modified cosmologies. The shaded curve 
represents the confidence region of the Hubble parameter for the Yukawa 
cosmology at $3\sigma$ CL. Additionally, we depict the OHD sample for further 
comparison. Both curves and the confidence region were obtained with the results 
of our MCMC analysis described in Section \ref{Sec:Constraints} for the SNe 
Ia+OHD.}}}
\end{figure}

\begin{figure}[ht!]
    \centering
    \includegraphics[scale = 0.56]{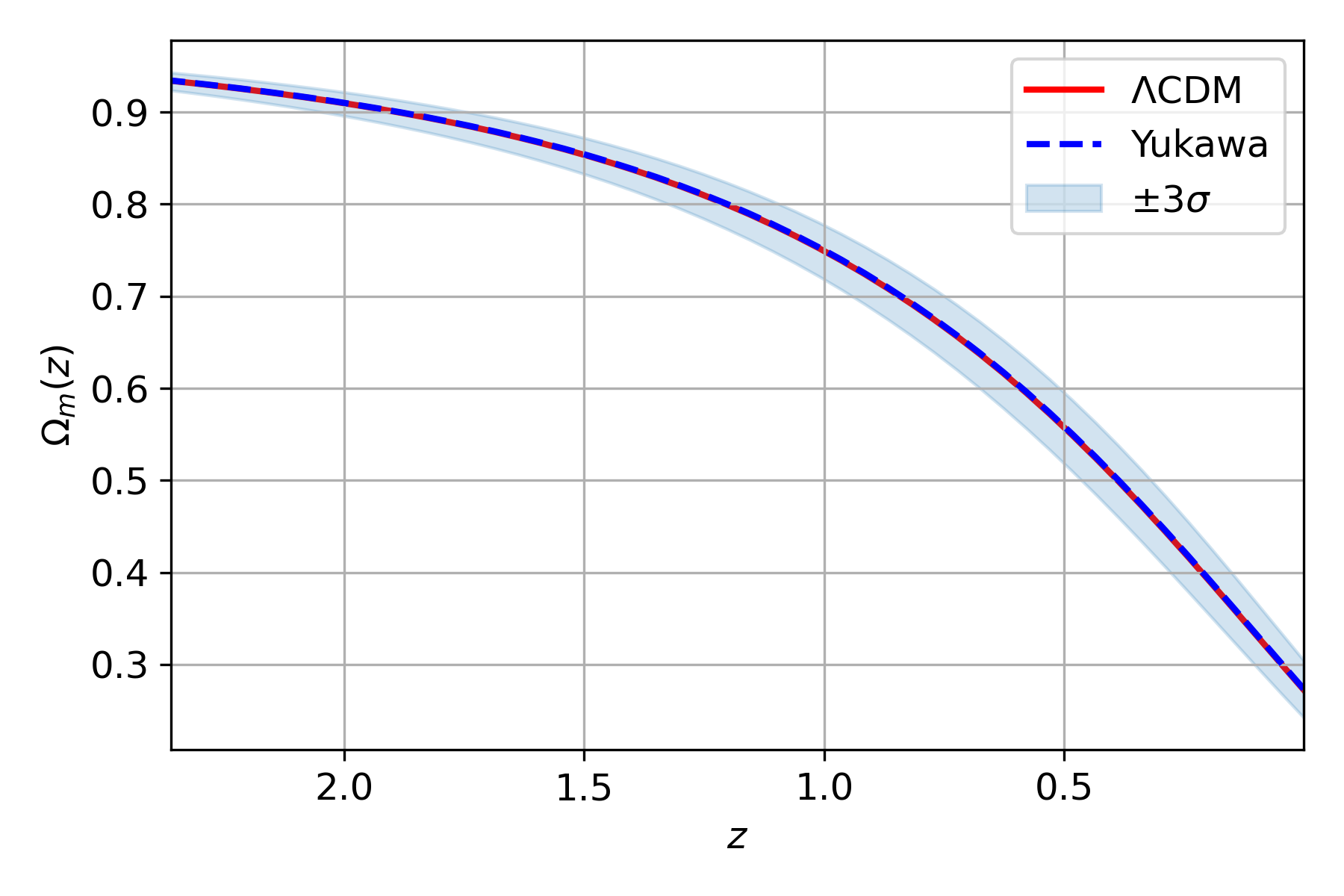}
    \caption{{\it{\label{fig:MYukawaplot} Matter density parameter as a function 
of redshift for the $\Lambda$CDM and Yukawa modified cosmologies. The shaded curve represents the confidence region of the Matter density parameter for the 
Yukawa cosmology at $3\sigma$ CL. Both curves and the confidence region were 
obtained with the results of our MCMC analysis described in Section 
\ref{Sec:Constraints} for the SNe Ia+OHD.}}}
\end{figure}

To establish the last point, we use the results of our MCMC analysis for SNe Ia+OHD to calculate the values of 
$\Omega_{B,0}^{\Lambda\text{CDM}}$, $\Omega_{DM,0}^{\Lambda\text{CDM}}$, and 
$\Omega_{\Lambda,0}^{\Lambda\text{CDM}}$  at $1\sigma$ 
CL from their definitions given by Eqs. 
\eqref{OB0def}, \eqref{ODMdef}, and \eqref{OLdef}, obtaining: 
$\Omega_{B,0}^{\Lambda\text{CDM}}=0.038\pm 0.003$, 
$\Omega_{DM,0}^{\Lambda\text{CDM}}=0.235\pm 0.008$, and 
$\Omega_{\Lambda,0}^{\Lambda\text{CDM}}=0.727\pm 0.011$. Note that 
$\Omega_{B,0}^{\Lambda\text{CDM}}+\Omega_{DM,0}^{\Lambda\text{CDM}}=0.273\pm 
0.011$, which match with the value of $\Omega_{m,0}$ obtained in the 
$\Lambda$CDM model. Therefore,   Yukawa   cosmology can mimic the 
late-time $\Lambda$CDM model, as we can see from Fig. \ref{fig:HYukawaplot}, 
where we depict the Hubble parameter as a function of redshift $z$ for the 
Yukawa   and $\Lambda$CDM cosmologies, given respectively by 
Eqs. \eqref{YukawaLCDMLike} and \eqref{HLCDM}. Furthermore, in 
Fig. \ref{fig:MYukawaplot} we depict  the matter density parameter $\Omega_{m}$ 
as a function of redshift for both models, were 
$\Omega_{m}=\Omega_{m,0}(1+z)^{3}/E^{2}$. Both figures were 
obtained with the results of our MCMC analysis described in Section 
\ref{Sec:Constraints} for the joint analysis.

\begin{table*}[ht!]
    \centering
        \begin{tabularx}{\textwidth}{YYYYY}
        \hline\hline
        \multirow{2}{*}{Data} &\multicolumn{3}{c}{Constraint} & \multirow{2}{*}{Reference} \\
        \cline{2-4}
         & $\alpha$ & $\lambda\;\text{[Mpc]}$ & $m_{g}\;\text{[GeV]}$ & \\
        \hline \\
        Galaxy Clusters & $\cdots$ & $>0.324$ & $<10^{-38}$ & A. S. Goldhaber \textit{et al.} (1974) \cite{Goldhaber:1974wg} \\
        Solar System & $\cdots$ & $>9.074\times 10^{-8}$ & $<7.2\times 10^{-32}$ & C. Talmadge \textit{et al.} (1998) \cite{Talmadge:1988qz} \\
        Weak Lensing & $\cdots$ & $>97.22$ & $<6\times 10^{-41}$ & S. R. Choudhury \textit{et al.} (2004) \cite{Choudhury:2002pu} \\
        BH  binaries & $\cdots$ & $> 9.722\times 10^{-4}$ & $<4 \times 10^{-35}$ & E. Berti \textit{et al.} (2011) \cite{Berti:2011jz}\\
        Local Group & $<0.581$ & $\cdots$ & $\left(3.265 _{-1 . 830}^{+1 . 830} \right)\times 10^{-35}$ & D. Benisty \textit{et al.} (2023) \cite{Benisty:2023ofi} \\
        SNe Ia+OHD & $0.416_{-0.326}^{+1.137}$ & $2693_{-1262}^{+1191}$ & $\left(2.374_{-0.728}^{+2.095}\right)\times 10^{-42}$ & Present paper \\ 
       \hline\hline
    \end{tabularx}
    \caption{Estimations of the coupling constant $\alpha$, the wavelength parameter ($\lambda$), and graviton mass ($m_g$) for the Yukawa potential, compared with our best-fit values using SNe Ia+OHD.}
    \label{tab:Comparison-Yukawa_parameters}
\end{table*}
\begin{figure*}
    \centering
    \includegraphics[scale=0.65]{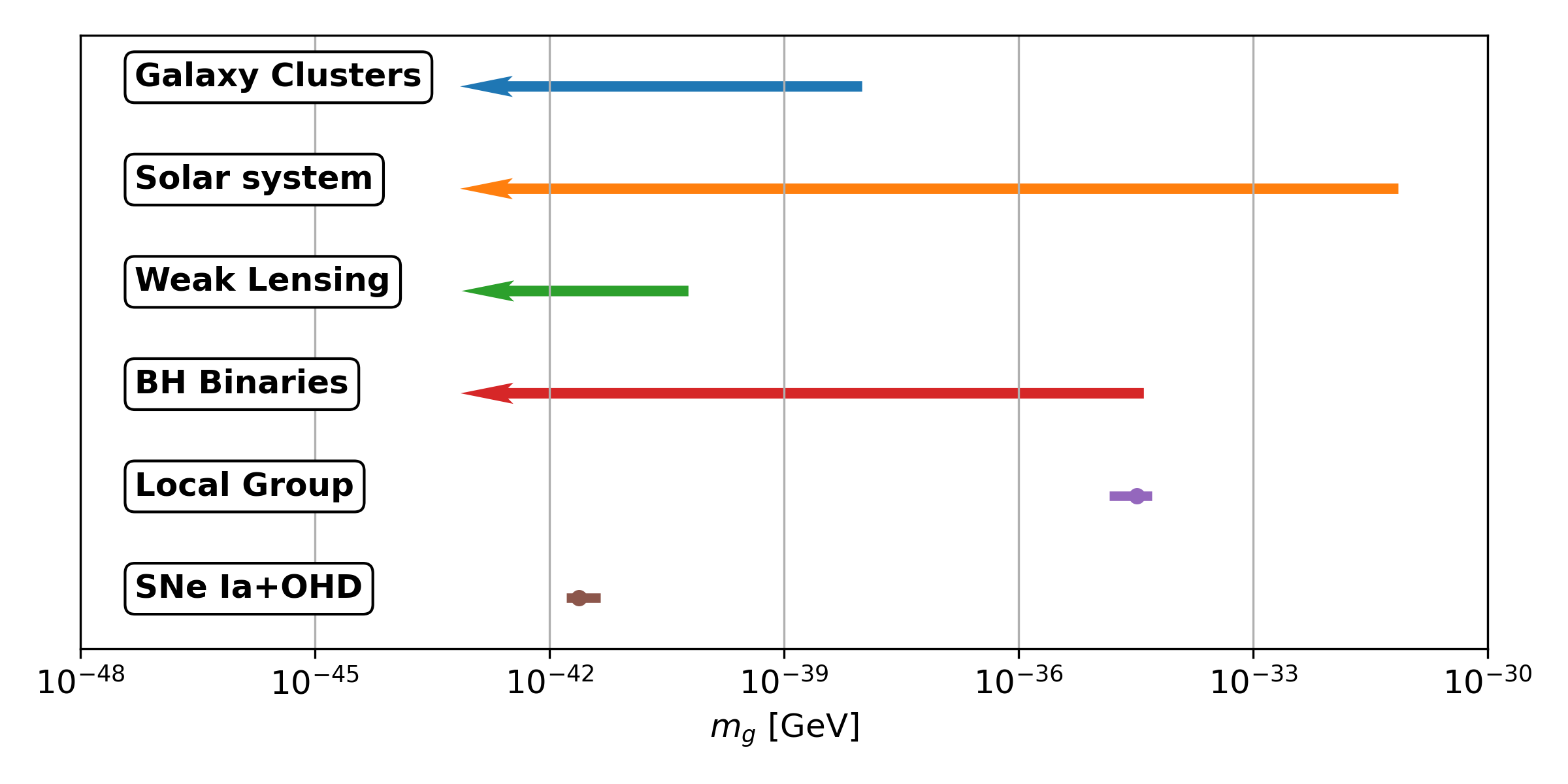}
    \caption{Upper bounds on the graviton mass ($m_g$) for the Yukawa potential, compared with our best-fit values using SNe Ia+OHD. For representation purposes, the $x$-axis is presented on a logarithmic scale.}
    \label{fig:mg}
\end{figure*}

The joint analysis has revealed the best-fits values for Yukawa parameters at $1\sigma$ CL. These are $\lambda=2693_{-1262}^{+1191}\, \text{Mpc}$, $\alpha=0.416_{-0.326}^{+1.137}$, and $m_{g}:= \hbar/(c\lambda)=\left(4.233_{-1.298}^{+3.735}\right)\times 10^{-69}\, \text{kg} \equiv \left(2.374_{-0.728}^{+2.095}\right)\times 10^{-42}\, \text{GeV}$.

Other studies have estimated the graviton mass differently. For instance, A. S. Goldhaber \textit{et al.} (1974) \cite{Goldhaber:1974wg}, using data from Galaxy Clusters, obtained $\lambda>0.324$ and $m_g<10^{-38}\;\text{GeV}$. C. Talmadge \textit{et al.} (1998) \cite{Talmadge:1988qz}, using data from Solar System measurements, obtained $\lambda >9.074\times 10^{-8}$ and $m_g<7.2\times 10^{-32} \;\text{GeV}$. Moreover, S. R. Choudhury \textit{et al.} (2004) \cite{Choudhury:2002pu}, using Weak Lensing measurements, obtains $\lambda>97.22$ and $m_g<6\times 10^{-41}\;\text{GeV}$. E. Berti \textit{et al.} (2011) \cite{Berti:2011jz}, observed multiple inspiralling black holes to determine the bounds that space-based detectors could place on the graviton Compton wavelength and graviton mass, leading to $\lambda> 9.722\times 10^{-4}$ and  $m_g<4 \times 10^{-35} \;\text{GeV}$. Finally, D. Benisty \textit{et al.} (2023) \cite{Benisty:2023ofi} obtain a value of $\alpha<0.581$ and $m_{g}<5.095\times 10^{-35}$ with Local Group estimations. We have compared our best-fit values for $\alpha$, $\lambda$, and $m_g$ at $1\sigma$ CL with previous references in Table \ref{tab:Comparison-Yukawa_parameters}.

Our results are consistent with more stringent Yukawa constraints on $m_g$ \cite{Choudhury:2002pu}. For instance, our findings show that the graviton mass is less than $6\times 10^{-41} \;\text{GeV}$ and $\lambda$ is greater than $97.22 \;\text{Mpc}$ obtained for Weak Lensing measurements \cite{Choudhury:2002pu}, and $\alpha$ is less than the value of $0.581$ obtained for the Local Group estimations \cite{Benisty:2023ofi}. Also, we provide Diagram \ref{fig:mg} for upper bounds on the graviton mass ($m_g$) for the Yukawa potential, which we compared with our best-fit values using SNe Ia+OHD. Our research has revealed that the graviton's mass is lower than previously estimated. This discovery significantly contributes to the scientific community and complements other significant findings related to Yukawa constraints from various systems  \cite{Goldhaber:1974wg, Talmadge:1988qz, Choudhury:2002pu, Berti:2011jz, deRham:2016nuf, Shao:2020fka, Benisty:2022txp, Benisty:2023ofi, Benisty:2023qcv}.

\section{Relating Yukawa cosmology and black hole shadows }
\label{IV}

 In this section, we will present a connection between the dark matter/dark energy 
densities and the angular radius of the black hole shadow. We will closely 
follow the approach developed in \cite{Tsupko:2019pzg, 
Escamilla-Rivera:2022mkc}, in which one can employ the standard definition of 
the luminosity distance for a flat $\Lambda$CDM model as \citep{Tsupko:2019pzg, 
Escamilla-Rivera:2022mkc}
\begin{equation} \label{eq:luminosity}
     d_L(z) = (1+z) c I(z)/H_0,
\end{equation}
where  $c$ is the speed of light, $H_0$ is the Hubble constant.   The quantity 
$I(z)$ is given in terms of the integral
\begin{equation} \label{eq:cosmointegral}
 I(z) = \int_{0}^{z} \left(\Omega_{m,0}^{\Lambda\text{CDM}} (1+\Tilde{z})^3 + 
\Omega_{\Lambda,0}^{\Lambda\text{CDM}}\right)^{-1/2} ~d\Tilde{z},
\end{equation}
with 
$\Omega_{m,0}^{\Lambda\text{CDM}}=\Omega_{B,0}^{\Lambda\text{CDM}}
+\Omega_{DM,0}^ {\Lambda\text{CDM}}$, that provides 
the present values of the critical density parameters for matter and a dark 
energy component, respectively. 
On the other hand, one can define the luminosity distance, which  is related to 
the angular diameter distance $d_A(z)$ in terms of the equation 
\citep{Tsupko:2019pzg, Escamilla-Rivera:2022mkc}:
\begin{equation} \label{eq:reciprocity}
    d_L(z) = (1+z)^2 d_A(z).
\end{equation}

To examine the connection with black holes shadows, let us recall 
that, by definition, the observed angular diameter of an object (say, a black 
hole) is given by $\theta=R/d_A$, with $R$ is some proper diameter of the 
object. Hence, we can extract information about the cosmological parameters, 
having measured one of these distances at a particular redshift $z$. Assuming 
Schwarzschild black holes, as a first approximation, an observer located far away 
from the black hole can construct the shadow image in the center with an angular 
radius  
\begin{equation} \label{angular shadow}
   \hat{ \alpha}_{\rm SH}(z) =  {R_{\rm SH}}/{d_A(z)},
\end{equation}
where $R_{\rm SH}$ is the shadow radius  and 
$M_{\rm BH} $ is the mass of the supermassive black hole. As it was pointed out 
in \citep{Escamilla-Rivera:2022mkc}, the above equation for the angular radius 
of the black hole shadow is valid only when the radial coordinate is large 
enough in comparison with the size of the black hole shadow radius 
$3\sqrt{3}GM_{\rm BH} /c^2$.
If we now  combine Eqs.~(\ref{eq:luminosity})--(\ref{eq:reciprocity}), we can 
obtain \citet{Tsupko:2019pzg,Escamilla-Rivera:2022mkc}
\begin{equation}\label{eq:luminosity shadow}
    \hat{\alpha}_{\rm SH} = \frac{R_{\rm SH}}{(1+z)} \frac{c}{H_0}I(z).
\end{equation}

In the present work,  we are interested in the  low-redshift limit, hence 
utilizing Eqs.~(\ref{eq:cosmointegral}) and (\ref{eq:luminosity shadow})  we 
can obtain \citep{Tsupko:2019pzg, Escamilla-Rivera:2022mkc}
\begin{equation}\label{eq:lowred}
\hat{\alpha}_{\rm SH}= {R_{\rm SH} H_0}/{(cz)}, 
\end{equation}
implying that we can estimate the angular radius of the black hole if we know 
the Hubble constant, the redshift $z$ of the black hole, along with its mass. 
Since we have explicit expressions of the dark matter/dark energy parameters in 
terms of $H_0$ (see Eqs. \eqref{eqDM} and \eqref{Eq.51}), we can solve for 
$H_0$ in Eq. \eqref{Eq.51} and directly relate the angular radius with the dark 
energy density parameter as
\begin{eqnarray}
    \hat{\alpha}_{\rm SH} =\frac{R_{\rm SH}}{z \lambda (1+\alpha)} 
\sqrt{\frac{\alpha}{ \Omega_{\Lambda,0}}},
\end{eqnarray}
or using the notation of \eqref{HLCDM}:
\begin{eqnarray}
    \hat{\alpha}_{\rm SH}  =\frac{R_{\rm SH}}{z \lambda} \sqrt{\frac{\alpha}{ 
(1+\alpha)\Omega_{\Lambda,0}^{\Lambda\text{CDM}}}}.
\end{eqnarray}
Similarly, we can express this relation in terms of the effective dark matter 
mass as  
\begin{eqnarray}\label{eq(56)}
    \hat{\alpha}_{\rm SH}  =\frac{R_{\rm SH}}{z \lambda (1+\alpha)} 
\frac{\sqrt{2 \alpha \Omega_{B,0}}}{\Omega_{D,0}},
\end{eqnarray}
or in terms of the notation  of \eqref{OB0def}, \eqref{ODMdef} and 
\eqref{OLdef}, as
\begin{eqnarray}
    \hat{\alpha}_{\rm SH} =\frac{R_{\rm SH}}{z \lambda \sqrt{1+\alpha}} \frac{\sqrt{2 \alpha \Omega_{B,0}^{\Lambda\text{CDM}}}}{\Omega_{D,0}^{\Lambda\text{CDM}}}.
\end{eqnarray}
In summary,   we can obtain the angular radius of black holes using 
dark matter and dark energy densities. Note that in this modified 
cosmological scenario, the shadow radius depends on the distance from the black 
hole to the observer. 
\subsection{Black hole solution}
Let us see how the spacetime geometry around the 
black hole us modified in this theory. The general solution in case of a 
static, spherically symmetric source reads
\begin{equation}
ds^2
= -f(r) dt^2 +\frac{ dr^2}{f(r)} +r^2(d\theta^2+\sin^2\theta d\phi^2).
\end{equation}
The energy density of the modified matter can be computed from 
$
    \rho(r)=\frac{1}{4\pi} \Delta \Phi(r)
$. In astrophysical scales we can set $l_0/r \to 0$, in that case using 
(\ref{eq11a}) we acquire 
\begin{eqnarray}
    \rho(r)=-\frac{M \alpha}{4 \pi r \lambda^2} e^{-\frac{r}{\lambda}}.
\end{eqnarray}
The negative sign reflects that the energy conditions are violated inside the black hole. On the other hand, we assume that Einstein field equation with a cosmological constant   holds in the sense that the effect of effective dark matter is encoded in the total energy-momentum part; namely, $G_{\mu \nu}+\Lambda g_{\mu \nu}=8\pi T_{\mu \nu}$.  Then, from the gravitational field equations for the $t-t$ component, we obtain 
    \begin{eqnarray}
        \frac{r f'(r)+f(r)-1}{r^2}+\Lambda-\frac{2 M \alpha}{ r \lambda^2} e^{-\frac{r}{\lambda}}=0,
    \end{eqnarray}
    yielding the solution 
    \begin{eqnarray}
        f(r)=1-\frac{2M}{r}-\frac{2M \alpha (r+\lambda) e^{-\frac{r}{\lambda}}}{r \lambda}-\frac{\Lambda r^2}{3}.
    \end{eqnarray}
The third term is due to the apparent dark matter effect, while the last term is the contribution due to the cosmological constant. We can perform a series expansion around $x=1/\lambda$, yielding
     \begin{eqnarray}
        f(r)=1-\frac{2M (1+\alpha)}{r}+\frac{M \alpha r}{\lambda^2}-\frac{\Lambda r^2}{3}+...
    \end{eqnarray}
It is, therefore, natural to define the true or the physical mass of the black hole to be $\mathcal{M} = M (1+\alpha)$ and write the solution in terms of the physical mass
 \begin{eqnarray}
        f(r)=1-\frac{2 \mathcal{M}}{r}+\frac{\mathcal{M} \alpha r}{(1+\alpha)\lambda^2}-\frac{\Lambda r^2}{3}+... \label{eq(60)}
    \end{eqnarray}
At this point, it is interesting to notice that the last equation can be linked to the  spherical black hole solutions in de Rham, Gabadadze and Tolley (dRGT) massive gravity \cite{Ghosh:2015cva}
 \begin{eqnarray}
        f(r)=1-\frac{2 \mathcal{M}}{r}-\frac{\Lambda r^2}{3}+\gamma \,r+\zeta,
    \end{eqnarray}
for $\zeta=0$ and $\gamma=\mathcal{M}\alpha/\lambda^2$. 
Further, we  can neglect the term $\mathcal{O}(\alpha/\lambda^2)$ in \eqref{eq(60)} and we obtain the Kottler spacetime, i.e., Schwarzschild black hole with a cosmological constant
 \begin{eqnarray}
        f(r)\simeq 1-\frac{2 \mathcal{M}}{r}-\frac{\Lambda r^2}{3}.
    \end{eqnarray}    
Following \citep{Vagnozzi:2022moj}, it is easy to show that the shadow radius for the metric \eqref{eq(60)} can be written as
\begin{equation}
   R_{\rm SH}=\frac{r_{\rm ph}}{ \sqrt{f(r_{\rm ph})}}\sqrt{1-\frac{2 GM_{\rm BH}}{c^2 r_O}+\frac{G M_{\rm BH} \alpha r_O}{c^2 (1+\alpha) \lambda^2}-\frac{1}{3}\Lambda r_O^2},
\end{equation}
where $r_O$ is the distance to the black hole. 
Hence, using the best-fit values of the previous section,
one can show that $\Lambda \simeq 10^{-52} \rm 
m^{-2}$, and we can identify the physical mass of the black hole with $\mathcal{M}=M_{\rm BH}$. For the Sgr A BH we can take $ M_{\rm BH}^{\rm Sgr A}=4 \times 10^6 M_{\rm Sun}$ with the distance $r_O=8.3 $ kpc. The change in the shadow radius compared to the Schwarzschild black hole is of the order $\delta R_{\rm SH}\sim 2\times 10^{-9}$ measured in black hole units. For the M87 black hole, we can take $ M_{\rm BH}^{\rm M87}=6.6 \times 10^9 M_{\rm Sun}$, along with the distance $r_O=16.8 $ Mpc, which compared to the Schwarzschild black hole changes by $|\delta R_{\rm SH}|\sim 2 \times 10^{-5}$. In other words, such a change is outside the scope of the present technology. We can approximate the shadow 
radius in both cases to be $R_{\rm SH} \simeq 3 \sqrt{3} GM_{\rm BH}/c^2$. 

Now,  it is well known that the real part of quasinormal modes 
$\omega_{\Re}$ in the eikonal limit is related to the shadow radius of BHs via  
$\omega_{\Re} =\lim_{l \gg 1}\frac{1}{R_{\rm SH}}\left(l+1/2\right)$ 
\citep{Jusufi:2019ltj, Jusufi:2020dhz, Jusufi:2020mmy, Cuadros-Melgar:2020kqn, 
Liu:2020ola,Zhao:2023uam} where $l$ is the angular node number. This 
correspondence is achieved based on the geometric-optics correspondence between 
the parameters of a quasinormal mode and the conserved quantities along 
geodesics. This connection allows the testing of gravitational waves with the 
next-generation Event Horizon Telescope \citep{Yang:2021zqy,Franchini:2023eda}. However, here we see that the frequency of the 
quasinormal modes emitted by a perturbed black hole in the eikonal limit will 
depend on the effect of cosmological constant and apparent dark matter. In 
particular, we obtain the following relation $
      \hat{\alpha}_{\rm SH} =\lim_{l\gg 1}\frac{(l+1/2)} {\omega_{\Re}\,z \lambda} 
\sqrt{\alpha/ (1+\alpha)\Omega_{\Lambda,0}^{\Lambda\text{CDM}}}$,  
 and 
$
    \hat{\alpha}_{\rm SH} =\lim_{l\gg 1}\frac{(l+1/2)}{\omega_{\Re}\,z \lambda 
\sqrt{1+\alpha}}  \sqrt{2 \alpha 
\Omega_{B,0}^{\Lambda\text{CDM}}}/\Omega_{D,0}^{\Lambda\text{CDM}}$,
respectively. These relations are valid in specific conditions, i.e.  the 
eikonal regime and the low-redshift limit. In what follows,  we will apply Eq. 
\eqref{eq(56)} to compute the angular radius, assuming  known  black hole 
mass and Yukawa parameters. 

\begin{itemize}
    \item Case I: Sgr A supermassive BH
    \end{itemize}
Using the best-fit values for $\lambda$ and $\alpha$ along  with the black hole 
mass for SgrA, we can estimate the angular radius:
\begin{eqnarray}
    \hat{\alpha}_{\rm SH}^{\rm Sgr A} =\frac{3 \sqrt{3}\,  G M_{\rm BH}^{\rm Sgr 
A}}{c^2\,z \lambda} \sqrt{\frac{\alpha}{ 
(1+\alpha)\Omega_{\Lambda,0}^{\Lambda\text{CDM}}}}\simeq 26.2 \rm \mu as,
\end{eqnarray}
where we have used $ M_{\rm BH}^{\rm Sgr A}=4 \times 10^6  M_{\rm Sun}$, 
$z=0.1895 \times 10^{-5}$, $\Omega_{\Lambda,0}^{\Lambda\text{CDM}}\sim 0.7$, 
$\lambda \simeq 2693$ [Mpc] and $\alpha \simeq 0.416$. Since we showed that the change in the shadow radius is $\delta R_{\rm SH}\sim 2\times 10^{-9}$, here we have approximated the shadow radius to $R_{\rm SH}\simeq 3 \sqrt{3} G M_{\rm BH}^{\rm Sgr A}/c^2$.
\begin{itemize}
     \item Case II: M87 supermassive BH
     \end{itemize}
     For the case of M87, we obtain:
     \begin{eqnarray}
    \hat{\alpha}_{\rm SH}^{\rm M87} =\frac{3 \sqrt{3}\, G M_{\rm BH}^{\rm 
M87}}{c^2\,z  \lambda} \sqrt{\frac{\alpha}{ 
(1+\alpha)\Omega_{\Lambda,0}^{\Lambda\text{CDM}}}}\simeq 19.13\rm \mu as,
\end{eqnarray}
where we have used $ M_{\rm BH}^{\rm M87}=6.6 \times 10^9 M_{\rm Sun}$, $z=0.428 
\times 10^{-2}$, along with $\Omega_{\Lambda,0}^{\Lambda\text{CDM}}\sim 0.7$, 
$\lambda \simeq 2693$ [Mpc] and $\alpha \simeq 0.416$. Again, the shadow radius is approximated to be $R_{\rm SH}\simeq 3 \sqrt{3} G M_{\rm BH}^{\rm Sgr A}/c^2$.  These values are 
consistent with those reported by the EHT \citep{EventHorizonTelescope:2019dse, 
EventHorizonTelescope:2020qrl, EventHorizonTelescope:2021srq, 
EventHorizonTelescope:2022wkp, EventHorizonTelescope:2022apq, 
EventHorizonTelescope:2022wok}. 

\section{Conclusions}
\label{conclusions}

In the present work,   we extracted    observational constraints on the Yukawa 
cosmological model. In this scenario,  dark matter  appears effectively, and a relation exists between  dark matter, dark energy, and 
baryonic matter. In particular, the   effective    dark matter is 
attributed to  the long-range force between the baryonic matter particles.   
Such a Yukawa-like gravitational potential modifies Newton's law of gravity in 
large-scale structures. It is characterized by the coupling parameter $\alpha$ 
and has a graviton with non-zero mass (which is inversely related to the 
wavelength parameter $\lambda$). We used   SNe Ia and 
OHD observational data, and we found within $1\sigma$ CL the best-fit parameter  
$\lambda=2693_{-1262}^{+1191}\, \rm Mpc$ and $\alpha=0.416_{-0.326}^{+1.137}$, 
respectively. With these values, we acquire the value $m_g =\left(4.233_{-1.298}^{+3.735}\right)\times 10^{-69}\, \text{kg}$, or, equivalently, $m_{g}=\left(2.374_{-0.728}^{+2.095}\right)\times 10^{-42}\, \text{GeV}$, for the graviton mass, complementing other significant findings related to Yukawa constraints from different systems \cite{Goldhaber:1974wg, Talmadge:1988qz, Choudhury:2002pu, Berti:2011jz, deRham:2016nuf, Shao:2020fka, Benisty:2022txp, Benisty:2023ofi, Benisty:2023qcv}.

Additionally, we found a black hole solution and a relation between the dark matter/dark energy density 
parameters and the angular radius of black hole shadows. These equations allow 
us to constrain the graviton mass directly from the EHT results for Sgr A and 
M87 supermassive black holes. We can further use  the following Gaussian 
likelihoods $\mathcal{L}_{\rm Shadow}\propto\exp{(-\chi_{\rm Shadow}^{2}}/2)$, 
where 
$\chi_{\text{shadow}}^{2}=\sum_{i=1}[\hat{\alpha}_{\rm SH}^{\rm observed}-\hat{\alpha}_{\rm SH}^{\rm theory}/\sigma_{\hat{\alpha, i}}]^{2},
$ and the modify the Gaussian  likelihood for the joint analysis SNe 
Ia+OHD+Shadow as 
$\mathcal{L}_{\text{joint}}=\mathcal{L}_{\text{SNe}}+\mathcal{L}_{\text{OHD}}
+\mathcal{L}_{\text{shadow}}$. We will consider and explore  this 
possibility in a separate project.

Our research combines two systems from varying scales, successfully integrating cosmological constraints on black hole shadows. That establishes a multi-messenger constraint on gravity, modelled explicitly as the Yukawa potential. 
Our application mainly focuses on constraints related to late-time cosmology. However, constraints from higher redshifts, such as those derived from Cosmic Microwave Background (CMB) and Big Bang Nucleosynthesis (BBN), could effectively restrict the Yukawa parameters in future research.

\section*{CRediT authorship contribution statement}
\textbf{Esteban González:} Conceptualization, Methodology, Software, Validation, Formal analysis, Investigation, Writing – review \& editing. \textbf{Kimet Jusufi:} Conceptualization, Methodology, Formal analysis, Investigation, Writing – original draft, Writing – review \& editing, Supervision, Project administration. \textbf{Genly Leon:} Conceptualization, Writing – review \& editing, Visualization, Investigation, Formal analysis, Funding acquisition, Supervision, Project administration. \textbf{Emmanuel N. Saridakis:} Conceptualization, Formal analysis, Investigation, Writing – review \& editing, Supervision.

\section*{Declaration of competing interest}
The authors declare that they have no known competing financial interests or personal relationships that could have appeared to influence the work reported in this paper.

\section*{Acknowledgments}
E.G. thanks to Vicerrectoría de Investigación y Desarrollo Tecnológico (VRIDT) at Universidad Católica del Norte (UCN) by the scientific support of Núcleo de Investigación No. 7 UCN-VRIDT 076/2020, Núcleo de Modelación y Simulación Científica (NMSC). G.L. was funded by VRIDT at UCN through Resolución VRIDT No. 040/2022, Resolución VRIDT No. 054/2022, Resolución VRIDT No. 026/2023 and Resolución VRIDT No. 027/2023. E.G. and G.L thank the support of Núcleo de Investigación Geometría Diferencial y Aplicaciones, Resolución VRIDT No. 096/2022.

\section*{Data Availability}
The data underlying this article were cited in Section \ref{Sec:Constraints}.

\bibliographystyle{apsrev4-1}

\bibliography{main.bib}

\end{document}